# Universal Non-Volatile Resistive Switching Behavior in 2D Metal Dichalcogenides Featuring Unique Conductive-Point Random Access Memory Effect

Xiaohan Wu[1], Ruijing Ge[1], Yuqian Gu[1], Emmanuel Okogbue[2], Jianping Shi[3], Abhay Shivayogimath[3], Peter Bøggild[4], Timothy J. Booth[4], Yanfeng Zhang[3], Yeonwoong Jung[2], Jack C. Lee[1] and Deji Akinwande[1]

[1]Microelectronics Research Center, University of Texas at Austin, Austin, TX 78758. deji@ece.utexas.edu
[2]Nanoscience Technology Center, University of Central Florida, Orlando, FL, 32826
[3]Department of Materials Science and Engineering, College of Engineering, Peking University, Beijing 100871, China
[4]Department of Physics, Technical University of Denmark, Lyngby 2800 Kgs, Denmark

**Abstract**

Two-dimensional materials have been discovered to exhibit non-volatile resistive switching (NVRS) phenomenon. In our work, we reported the universal NVRS behavior in a dozen metal dichalcogenides, featuring low switching voltage, large on/off ratio, fast switching speed and forming free characteristics. A unique conductive-point random access memory (CPRAM) effect is used to explain the switching mechanisms, supported by experimental results from current-sweep measurements.
(Keywords: 2D materials, non-volatile memory, and resistive switching)

## Introduction

Two-dimensional (2D) materials have drawn much attention as a promising candidate in the next-generation electron devices, optoelectronics and bioelectronics [1]. Over the last few years, various 2D materials, including graphene oxide, solution-processed transitional metal dichalcogenides (TMDs), degraded black phosphorus and multilayer hexagonal boron nitride (h-BN), have been reported to exhibit non-volatile resistance switching (NVRS) phenomenon, in which the resistance can be reversibly switched between a high resistance state (HRS) and a low resistance state (LRS) through external electrical bias and maintained without power supply [2-3].

Recently, we reported that NVRS phenomenon is accessible in a variety of single-layer TMDs and single-layer h-BN in vertical MIM structure [4-7]. Compared with other 2D material-based NVRS devices, the single-layer h-BN atomic sheet has only one atomic layer and ~0.33 nm in thickness, which is the thinnest active layer in non-volatile resistance memory. Here, we expanded the collection of TMDs with NVRS characteristics. 12 different 2D TMD materials in the form of $MX_2$ (M=metal, e.g. Mo, W, Re, Sn or Pt; X=chalcogen, e.g. S, Se or Te) have been investigated, alluding to a universal resistive switching characteristic in TMD atomic sheets. The switching behavior can be explained by a unique "conductive-point" model based on metal adsorption into intrinsic vacancies in TMDs (a point effect in contrast to filamentary "conductive-bridge" model in traditional bulk materials). Current sweeping measurements have been performed on these 2D devices and provide additional insights into the switching behavior. Our work in a broad collection of TMD atomic sheets suggests that a universal conductive-point random access memory (CPRAM) effect may be realizable in hundreds of atomically-thin 2D materials that might advance diverse applications including high-density neuromorphic computing, non-volatile memory fabrics, and zero-power RF switches.

## Device Fabrication and Characterization

The devices used in this work are based on monolayer, few-layer or heterostructure TMD materials which were mostly synthesized by chemical vapor deposition (CVD). The bottom electrodes (BE) were patterned by e-beam lithography and deposited by e-beam evaporation on either a SiO2/Si or diamond substrate. The diamond substrate, with a high thermal conductivity (~1000 W m$^{-1}$ K$^{-1}$), can dissipate the excessive heat effectively, thus preventing narrow electrode metal lines from Joule heating in small-area device integration. The TMDs were then transferred onto the target substrate with BE on it. The top electrodes were prepared using the same fabrication process as BE. The schematic of the fabricated TMD-based NVRS device is shown in Fig. 1.

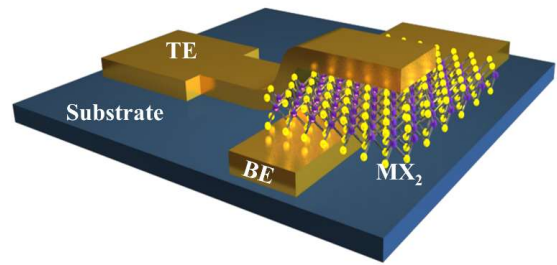

Fig. 1. The schematic of a crossbar device in vertical metal-insulator-metal (MIM) structure with transition metal dichalcogenides ($MX_2$) as the active layer.

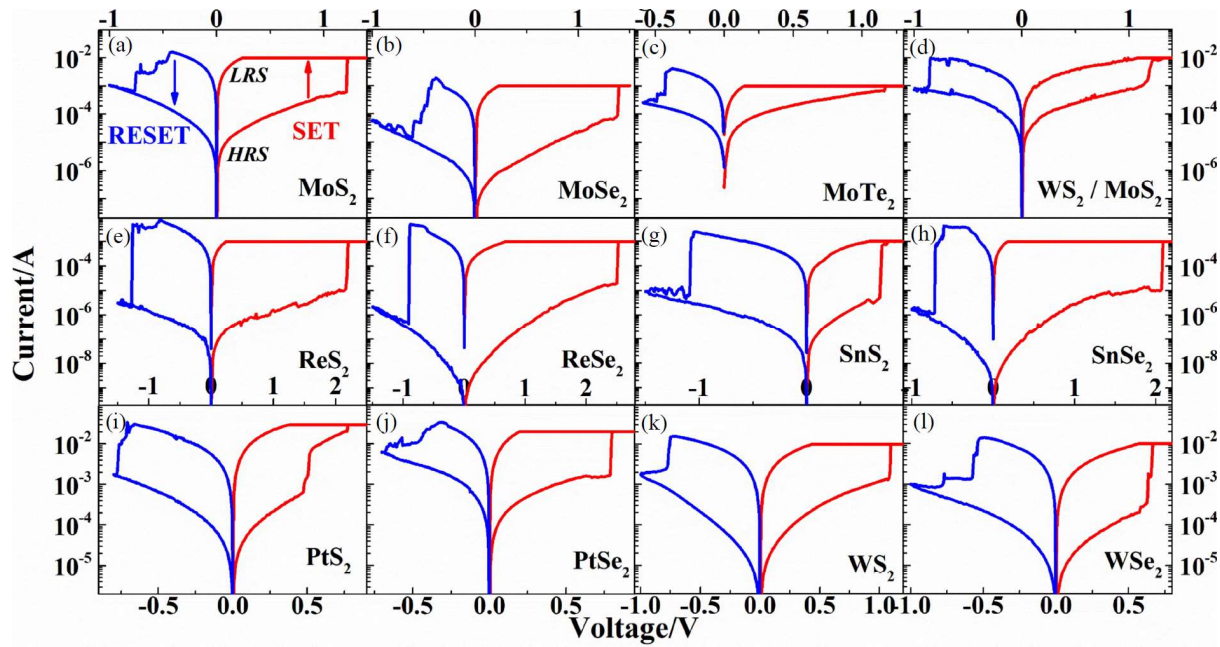

Fig. 2. Representative resistive switching I-V curves of 12 different 2D $MX_2$ materials including: (a) $MoS_2$, (b) $MoSe_2$, (c) $MoTe_2$, (d) $WS_2/MoS_2$ heterostructure, (e) $ReS_2$, (f) $ReSe_2$, (g) $SnS_2$, (h) $SnSe_2$, (i) $PtS_2$, (j) $PtSe_2$, (k) $WS_2$, and (l) $WSe_2$. All the $MX_2$ are monolayer materials except $MoTe_2$, $SnS_2$, $SnSe_2$ and $PtSe_2$ which are few layers due to the difficulty in obtaining high-quality and air-stable monolayers. The non-volatile resistive switching in these metal dichalcogenides indicates a universal phenomenon in conventional MIM configuration similar to the universality in transition metal oxides.

## Results and Discussions

Fig. 2 shows representative I-V curves of crossbar devices based on 12 different TMD materials. The NVRS phenomenon can be described with $MoS_2$ (Fig. 2a) as an example. The device commonly starts from a HRS. With voltage sweeping to ~ 1.2V, the current suddenly increases, indicating a transition from HRS to LRS (SET). For bipolar operation, a reverse voltage sweep is applied to reset the device back to HRS (RESET). This switching phenomenon can be observed in all the 12 TMD atomic sheets with similar characteristics. The difference among various TMDs is difficult to identify from the switching curves because of device-to-device and cycle-to cycle variations, a long-standing intrinsic challenge of NVRS devices due to the stochastic nature of the switching process. Future advances in 2D material growth, fabrication process optimization and testing protocol are expected to reduce material variability to figure out the differences of performance among these materials.

To illustrate the switching mechanisms in TMD-based NVRS devices, we proposed a "conductive-point" model based on metal adsorption into vacancies in TMD films. First, with external bias, metal ions/atoms can be dissociated from the electrodes. The metal ion(s) then adsorb into chalcogen defect(s). Electrons will transport through the metallic conductive point at the vacancy site(S, Se, or Te vacancies), resulting in a lower resistance state. The conductive-point model has been experimentally supported by atomic resolution STM measurement with the evidence of Au moving in and out from the vacancy site after SET and RESET process [8]. Compared with the conductive-bridge random access memory (CBRAM) effect in traditional bulk materials, the CPRAM effect indicates a unique feature in the ultra-thin 2D materials with atomic-level resistance control. The schematic of the conductive-point model with metal adsorption is illustrated in Fig. 3.

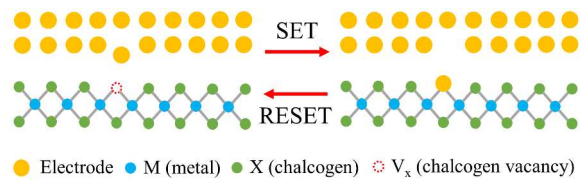

Fig. 3. Schematics of the conductive-point switching mechanism in $MX_2$ owing to electrode metal adsorption into chalcogen vacancy (BE not shown).

While voltage sweeping method is commonly utilized, current sweeping may reveal hidden properties. Fig. 4 displays the voltage-current relationship by current sweeping method for a MoS2 NVRS device. The transition starts from a HRS, followed by a gradual increase of both voltage and current. As the current reaches ~ $10^{-7}$A, the voltage suddenly drops while the conduction current remained the same, which means the resistance of the device is switched from a higher resistance state to a lower resistance state. Multiple subsequent voltage decreasing steps can be observed, leading to a transition from HRS to LRS. As the current sweeps to ~ 12 mA, the voltage abruptly increases, indicating a transition from LRS to HRS (RESET). Compared with RESET behavior using voltage sweep, a compliance voltage is required to avoid extremely high voltage across the device.

An observed correlation of transition steps has been established: For a device with single-step SET by current sweeping, the voltage-sweep RESET is also single-step (Fig. 5), while for a device with multiple steps during current-sweep SET, a multiple-step RESET can be observed by voltage sweeping. The transition steps in the switching curve can be attributed to the number of vacancies associated with the switching phenomenon, indicating multiple conductive points may exist in the resistive switching process in 2D-based NVRS devices.

## Conclusion

In conclusion, we reported a universal NVRS phenomenon in 12 different TMD materials. A unique atomic-level conductive-point model is proposed based on metal adsorption into chalcogen vacancy. Current sweeping method unveils the details hidden in the commonly used voltage-sweep curves, where the transition step number could be attributed to the number of defects/vacancies associated with resistive switching behavior. Our work provides a broad collection of 2D materials that exhibit NVRS phenomenon for the applications in neuromorphic computing and low-power RF switches.

## Acknowledgments

The authors acknowledge funding from a PECASE award, NSF #1809017, NSF NNCI #1542159. KETEP and MOTIE of the Republic of Korea, NKPs #2018YFA0703700, NSFC #51925201, DNRF103 (CoE CNG), EU Horizon2020 #696656 and #785219.

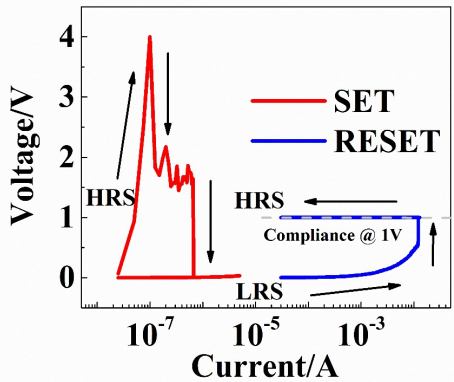

Fig. 4. The complete switching curve consisting of SET and RESET process by current sweeping.

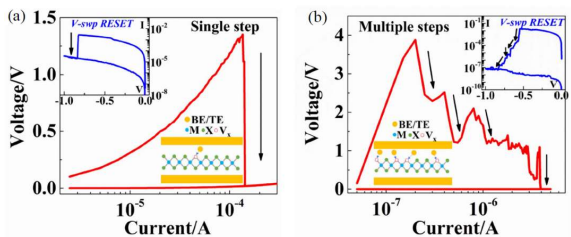

Fig. 5. A correlation of transition steps between voltage-sweep RESET and current-sweep SET. (a) single step, and (b) multiple steps, which is attributed to the number of vacancies associated with the switching phenomenon.

## References

[1] D. Akinwande et al., Nature, vol. 573, no. 7775, pp. 507-518, Sep 2019.
[2] Y. Shi et al., 2017 IEDM, San Francisco, CA, 2017, pp. 5.4.1-5.4.4.
[3] M. Wang et al., Nature Electronics, vol. 1, no. 2, pp. 130-136, Feb 2018.
[4] R. Ge et al., 2018 IEDM, San Francisco, CA, 2018, pp. 22.6.1-22.6.4.
[5] M. Kim et al., 2019 IEDM, San Francisco, CA, USA, 2019, pp. 9.5.1-9.5.4.
[6] X. Wu et al., Advanced Materials, vol. 31, no. 15, p. 7, Apr 2019.
[7] X. Wu et al., Nanotechnology, vol. 31, no. 46, Nov 2020, Art no. 465206
[8] S. M. Hus et al., "Single-defect memristor in MoS2 atomic-layer", arXiv, 2020.